\documentclass[]{emulateapj}
\usepackage{graphicx}
\usepackage{amsmath}
\usepackage{natbib}
\begin{document}
\title{A Distinctive Disk-Jet Coupling in the Seyfert-1 AGN NGC 4051}
\author{A. L. King\altaffilmark{1}, J. M. Miller\altaffilmark{1}, E. M.  Cackett\altaffilmark{2}, A. C. Fabian\altaffilmark{2}, S. Markoff\altaffilmark{3}, M. A. Nowak\altaffilmark{4}, M. Rupen\altaffilmark{5}, K. G\"ultekin\altaffilmark{1}, M. T. Reynolds\altaffilmark{1}}

\altaffiltext{1}{Department of Astronomy, University of Michigan, 500 Church Street, Ann Arbor, MI 48109, ashking@umich.edu}
\altaffiltext{2}{Institute of Astronomy, University of Cambridge, Madingley Road, Cambridge CB3 0HA, UK}
\altaffiltext{3}{Astronomical Institute 'Anton Pannekoek', University of Amsterdam, Science Park 904, 1098 XH, Amsterdam}
\altaffiltext{4}{\emph{Chandra} X-ray Science Center, Massachusetts Institute of Technology, NE80-6077, 77 Massachusetts Ave., Cambridge, MA 02139}
\altaffiltext{5}{Array Operations Center, National Radio Astronomy Observatory, 1003 Lopezville Road, Socorro, NM 87801}

\journalinfo{The Astrophysical Journal}
\submitted{Received 2010 July 28; accepted 2010 December 2}

\begin{abstract}
We report on the results of a simultaneous monitoring campaign employing eight \emph{Chandra} X-ray (0.5--10 keV) and six VLA/EVLA (8.4 GHz) radio observations of NGC 4051 over seven months. Evidence for compact jets is observed in the 8.4 GHz radio band; This builds on mounting evidence that jet production may be prevalent even in radio-quiet Seyferts. Assuming comparatively negligible local diffuse emission in the nucleus, the results also demonstrate an inverse correlation of $L_\mathrm{radio} \propto L_\mathrm{X-ray}^{-0.72 \pm 0. 04}$. If the A configuration is excluded in the case that diffuse emission plays a significant role, the relation is still $L_\mathrm{radio} \propto L_\mathrm{X-ray}^{-0.12 \pm 0. 05}$. Current research linking the mass of supermassive black holes and stellar-mass black holes in the ``low/hard" state to X-ray luminosities and radio luminosities suggest a ``fundamental plane of accretion onto black holes" that has a positive correlation of $L_\mathrm{radio} \propto L_\mathrm{X-ray}^{0.67\pm0.12}$. Our simultaneous results differ from this relation by more than 11 $\sigma$ (6 $\sigma$ excluding the A configuration), indicating that a separate mode of accretion and ejection may operate in this system. A review of the literature shows that the inverse correlation seen in NGC 4051 is seen in three other black hole systems, all of which accrete at near 10\% of their Eddington luminosity, perhaps suggesting a distinct mode of disk-jet coupling at high Eddington fractions. We discuss our results in the context of disk and jets in black holes and accretion across the black hole mass scale.

\end{abstract}

\section{Introduction}
With evidence of supermassive black holes (SMBH) lurking at the center of nearly all galaxies \citep{Richstone98}, it is pertinent to examine and understand their properties as well as their impacts. Studies have shown a critical relation between SMBHs and their host galaxies in the form of the $M$--$L$ and $M$--$\sigma$ relations. The $M$--$L$ relation between the mass of the SMBH and the luminosity of the bulge suggests an intrinsic link between the SMBH and the amount of mass in the bulge assuming a particular mass-to-light ratio \citep[e.g.][]{Magorrian98,Kormendy95,Gultekin09b}  The $M$--$\sigma$ relation between SMBH mass and the velocity dispersion of the host galaxy also implies a physical coupling between formation and growth of the black hole and its surroundings \citep[e.g][]{Ferrarese00,Gebhardt00}. The driving mechanism behind these couplings and the $M$--$\sigma$ relation especially, is thought to be the result of mergers that drive accretion onto the SMBH, which can quench star formation as energy released from the central engine drives the gas out \citep{DiMatteo05}.

In particular, the study of an accretion disk around a SMBH begins with observations of the extended blackbody spectrum emitted by the disk; a consequence of the radial dependence of the temperature associated with the accretion disk. In SMBH accretion disks, this spectrum is thought to peak in the UV and is associated with the ``Big Blue Bump" \citep{Elvis94}. Unfortunately, UV flux is extremely susceptible to scattering by dust and modeling to correct for this can induce large uncertainties. In X-rays, emission from inverse-Compton scattering, magnetic flares and magnetic reconnection events associated with the accretion disk are characterized well by a non-thermal power-law \citep[e.g.][]{McHardy04}. Accordingly, X-ray flux can be another proxy for observing accretion disks.

In accreting systems, as material migrates toward the center, a fraction is also ejected into outflows that have both radiative and mechanical influences on their environments. The exact physical nature has not yet been observationally determined, but outflows are seen in all types of accreting systems, i.e., proto-stellar objects \citep[e.g.][]{Mundt85}, neutron stars and stellar-mass black holes \citep[e.g.][]{Margon82}, and SMBHs \citep[e.g.][]{Cohen79}. These outflows can reach supersonic speeds when collimated into jets, eventually depositing significant energy into their surroundings \citep[e.g.][]{Cohen79,Fabian02,Allen06}. Material moving outward into their host galaxies also begins to cool via synchrotron radiation \citep[e.g.][]{Jones74}. Observed in the radio frequencies, this non-thermal process emitted in the core of the system is predicted to have a flat spectrum, independent of frequency \citep{Blandford79} making it a great observational tool for characterizing jet emission. 

Utilizing these two wavelength regimes, \cite{Merloni03}, \cite{Falcke04}, and \cite{Gultekin09} have all suggested a ``fundamental plane" of black hole activity connecting black hole mass, X-ray luminosity and radio luminosity. The plane spans over 9 orders of magnitude in mass, 12 orders of magnitude in radio luminosity and 13 orders of magnitude in X-ray luminosity \citep{Merloni03}. This plane suggests that accretion (traced by X-ray luminosity) and jet production (traced by radio luminosity) are fundamentally linked together. Although the exact coupling is not understood, this relation implies accretion must be driving jet production.

A known problem with the relation is the large scatter of the data about the plane,  \citep[$\sigma_{\mathrm{radio}}=0.88 {\rm dex}$][]{Gultekin09}. This scatter can be attributed to observational errors or the result of non-simultaneouty between the observations themselves. Measuring the X-ray and radio luminosities at different times may have sampled different fluctuations in the accretion rate in individual sources, driving them away from the relation. The time between X-ray and radio observations of \cite{Merloni03} and \cite{Gultekin09} are known to span a few years to a decade. 

To address these issues, and in order to examine disk-jet coupling at high mass accretion rates, we undertook a simultaneous X-ray and radio monitoring campaign of the Seyfert-1 AGN NGC 4051.  This galaxy is relatively nearby (z=0.002336), and the central black hole mass has been determined through reverberation mapping techniques \citep[$(1.91 \pm 0.78) \times 10^6 M_\odot$][$(1.73 \pm 0.55) \times 10^6 M_\odot$]{Peterson04,Denney09}. NGC 4051 is typically observed to accrete at approximately 5\% of its Eddington luminosity \citep{Peterson04}. The innermost orbital timescale of NGC 4051 is on the order of a few minutes to hours,  defined as $t_{dyn} \sim R/v_{\phi}$, where $R$ is the radius assumed to be only a few gravitational radii from the black hole and $v_{\phi}$ is the orbital velocity. While the viscous timescales are on order of days to weeks for typical parameters, defined as $t_{vis} \sim t_{dyn} \alpha^{-1} (H/R)^{-2}$, where $\alpha$  is the viscosity parameter in the standard $\alpha$-disk prescription \citep{Shakura73}, and $H$ is the scale-height of the disk. Variations in the accretion rate of NGC 4051 are thought to occur on these viscous timescales of a day to weeks. This is further supported by its highly variable spectrum, which is most notable in X-rays that vary up to a factor of 10 on weekly timescales \citep{Uttley99}. The variability seen in NGC 4051 was essential for our simultaneous monitoring campaign to probe different accretion rates. 

In this paper, we present the simultaneous X-ray and radio observations of NGC 4051, in effort to shed light on the implications of the fundamental plane of accretion onto black holes, and to explore jet-production in this Seyfert galaxy.

\section{Data Reduction and Anlaysis}
\subsection{X-ray}
The Advanced CCD Image Spectrometer (ACIS) on the \emph{Chandra} X-ray telescope was used to collect eight observations of NGC 4051 between 1 Jan 2009 (MJD 54848.2) and 31 Jul 2009 (MJD 55043.1). We list these observations of approximately 10 ks exposures in Table \ref{xrayobs}. X-ray observations are subject to photon pile-up. Photon pile-up is when multiple low-energy photons arrive at the same CCD pixel in a single frame and register identically to one higher energy photon. It reduces the number of soft, i.e. low energy, photons that are detected while increasing the hard, i.e. high energy, photon counts which can mislead analysis by making the spectrum appear harder than it actually is. To avoid photon pile-up, we used the mode with the minimal integration time, i.e. the Continuous Clocking mode, with integration times of only 2.85 ms. Such a fast integration time helps to prevent against multiple photons being counted as one higher energy photon, but does so at the expense of one spatial dimension. The resulting image is 1 $\times$ 1024 pixels with a resolution $<1''$, and an effective field of view of 8.3$'$ $\times$ 8.3$'$, with spatial resolution in only one dimension. 

We used \texttt{HEASOFT} version 6.7, \texttt{FTOOLS} version 6.7, \texttt{XSPEC} version 12.5.1, \texttt{CALDB} version 4.1 and \texttt{CIAO} version 4.1 in the data reduction and analysis of these images. The \texttt{CIAO} \texttt{psextract} routine was used to extract a spectrum and background for a point source as well as the companion auxiliary response functions (ARF) and response matrix functions (RMF). To do so, we used a circular extraction region of diameter $2''$ centered at 12h03m09.6s, $+44^\circ31'52\farcs5$, while another $2''$ diameter circular extraction region was used approximately $20''$ off-source for the background. This aperture size ensured encapsulation of the source, assuming it was a point source and not elongated, but not so large as to include effects from the background. 

We analyzed X-ray data between the energy range of 0.5--10.0 keV with a minimum of 10 counts per bin. A power-law component modeling the continuum was initially fit to the data. To check for pile-up in our source before further modeling was attempted, we also used an annulus with inner radius of $0\farcs9$ and outer radius of $1\farcs7$ to obtain an additional spectrum from each observation using \texttt{psextract}. Again, requiring a minimum of 10 counts per bin, we fit power-law models to the annular spectra for comparison. If pile-up was present, the spectra from the annuli would be softer than those extracted from the central region. In general, we found that the two spectra were equivalent, within errors, suggesting that any pile-up is minimal in all observations.

A description of the line-of-sight absorption is as follows and was included in all modeling.  The Galactic absorption was modeled as an effective H column density of $1.15$x$10^{20}$ cm$^{-2}$ \citep{Kalberla05} using \texttt{phabs}. Two absorption edges typical to Seyfert AGN \citep{Reynolds97} at K-shell rest energies of O VII and O VIII of 0.739 and 0.871 keV respectively, were modeled using \texttt{zedge}. These three rest-frame values were frozen, while individual normalizations were allowed to vary. 

\begin{figure*}[t]
\begin{center}
\includegraphics[angle=270,scale=.45, trim=0mm 0mm 10mm 0mm, clip]{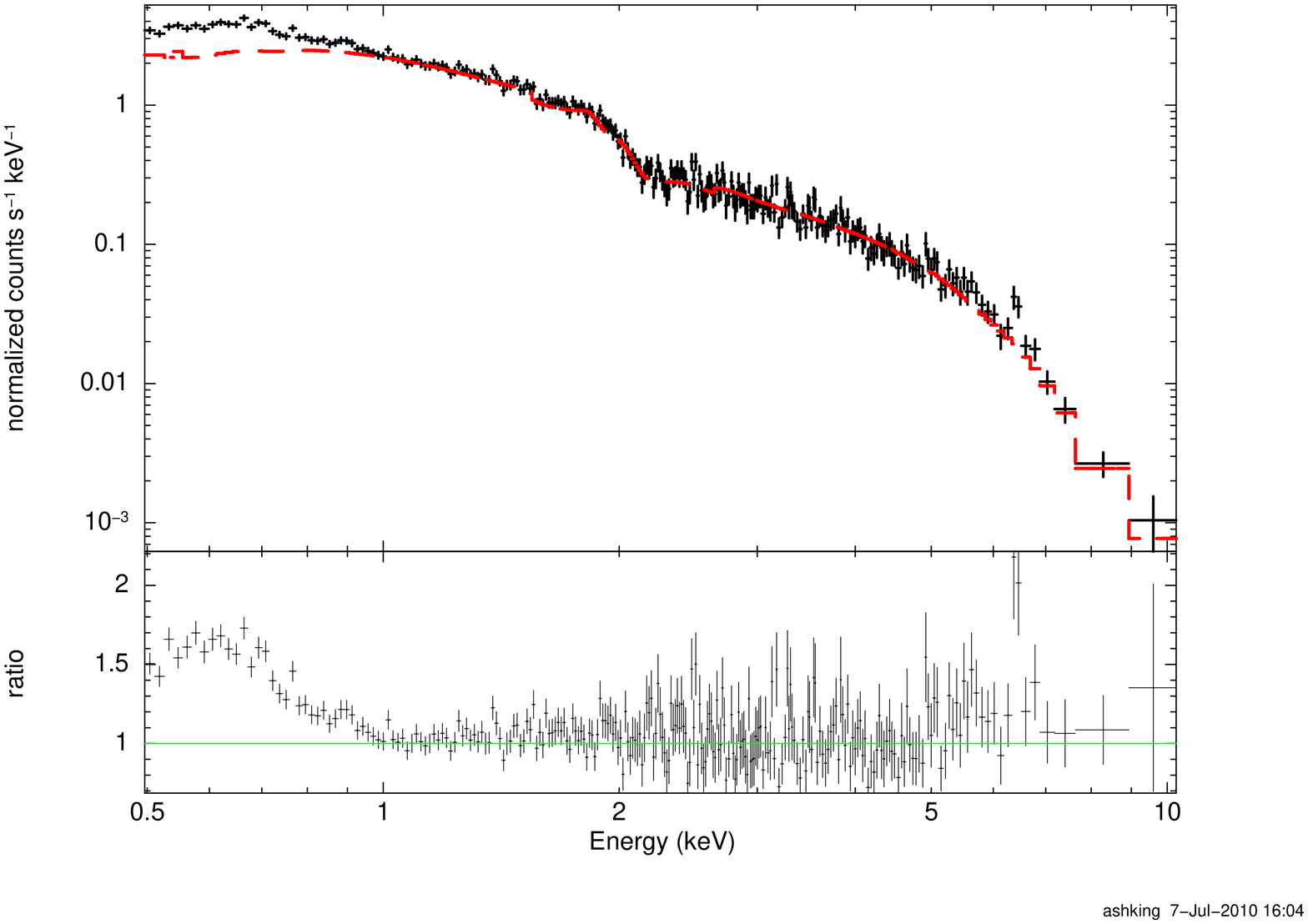}
\caption{\footnotesize{This is a sample spectrum from MJD 54943.1 modeled with just a power-law component, in red, as the \texttt{Xspec} model \texttt{phabs(po)*zedge*zedge}. The Galactic absorption was modeled as an effective H column density of $1.15$x$10^{20}$ cm$^{-2}$ \citep{Kalberla05} and the K-shell absorption edges of OVII and OVIII were frozen at their respective rest energies of 0.739 keV and 0.871 keV. We initially fit the spectrum from 2--10 keV and then extended the model to lower energies. The fit produced a $\chi^2/\nu$=1770/366. }}
\label{fig1}
\includegraphics[angle=270,scale=.45,trim=0mm 0mm 10mm 0mm,clip]{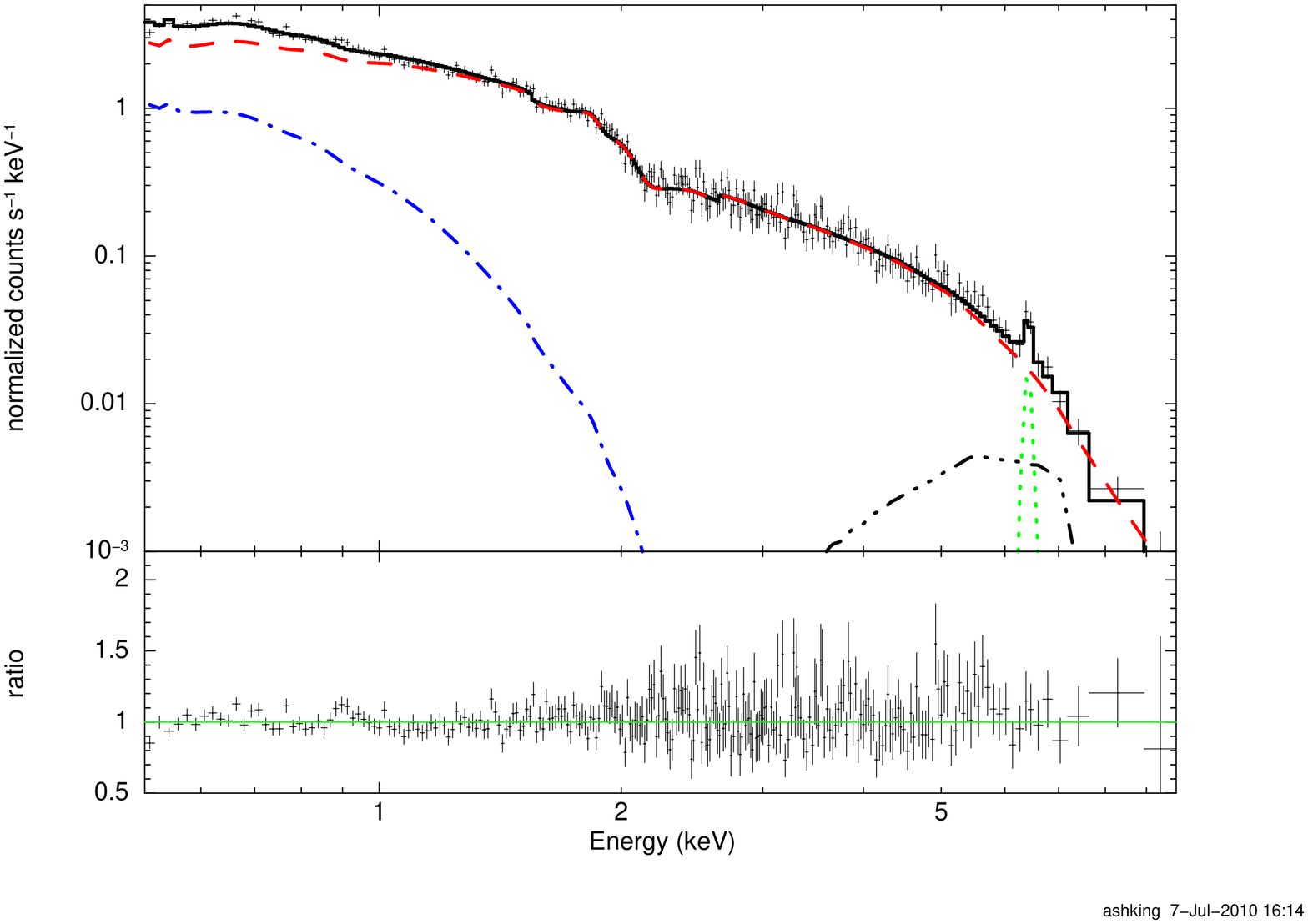}
\end{center}
\caption{\footnotesize{This is the same spectrum as Figure \ref{fig1} but now includes a power-law, disk blackbody, a narrow Fe K$\alpha$ line and a broad Fe K$\alpha$ line, modeled with \texttt{Xspec} model \texttt{phabs(po+diskbb+zgauss+laor)*zedge*zedge}. We used the same Galactic absorption and absorption edges as in Figure \ref{fig1}. The power-law is the red dashed line. The disk blackbody is the blue dot-dashed line. The narrow Fe K$\alpha$ line is the green dotted line. The broad Fe K$\alpha$ line is the black triple dotted-dashed line. Finally, the solid black line is the sum of all the components. The fit produced a $\chi^2/\nu$=368.0/353. }}
\label{fig2}
\end{figure*}

\begin{figure*}[t]
\begin{center}
\includegraphics[angle=90,scale=.45]{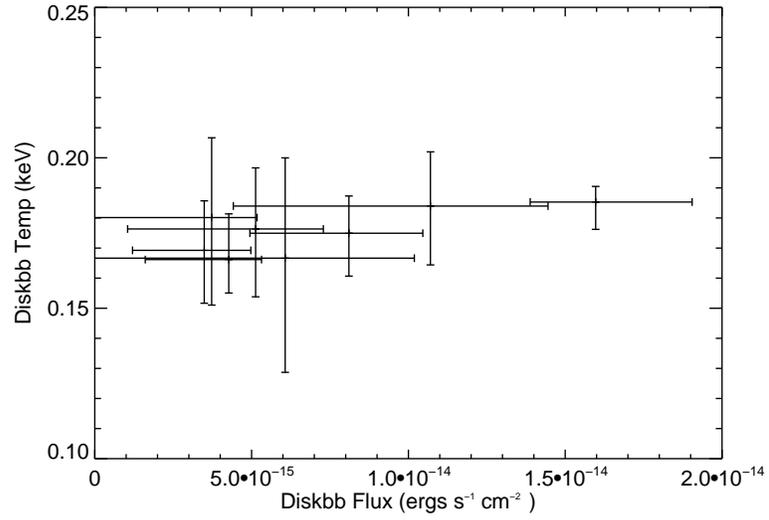}
\caption{\footnotesize{ This figure shows the comparison between the soft-excess disk blackbody temperature and its blackbody flux. A constant temperature independent of flux as seen here suggests a purely phenomenological interpretation of the disk blackbody component.}}
\label{bb}
\end{center}
\end{figure*}

Figure \ref{fig1} shows a sample spectrum fit with a simple power-law model. This attempt at fitting the data resulted in a poor fit for all spectra and in particular the spectrum presented in Figure \ref{fig1} had a $\chi^2/\nu$ of 1770/366. By looking at the ratio of the data to the model, one can see the obvious excess at low energies typical of Seyfert galaxies \citep{Reynolds97}. To better characterize the source flux, a disk blackbody component was added shown in Figure \ref{fig2}. Typically, it is thought that this soft excess is not the result of thermal blackbody component but of an atomic process that is not trivial to model \citep[e.g.][]{Gierlinski04,Crummy06}. Figure \ref{bb} plots the temperature and flux from the putative disk component. We find that the temperature does not vary with flux, suggesting it is not a blackbody thermal component. However, the addition of the disk blackbody does produce a formally better fit. 

In Figure \ref{fig2}, we not only include a power-law component and a disk blackbody but also an unresolved Gaussian that models a narrow Fe K$\alpha$ line and a broad Fe K$\alpha$ line. The fit is improved to $\chi^2/\nu$=368.3/353. It should be noted that the Fe K$\alpha$ lines are not detected at more than the $2\sigma$ level of confidence, except in the last two exposures. However, they were included in the modeling for completeness (see Table \ref{Feobs}). In Seyferts and AGN, Fe K-shell emission due to fluorescence and recombination is the most prominent of the X-ray emission lines \citep[e.g.][]{Miller07}. The shape and broadening of the Fe K$\alpha$ line, initially modeled by \cite{Fabian89} for a zero-spin Schwarzschild black hole and \cite{Laor91} for a maximally spinning black hole, is dependent on the spin of the black hole, the inner and outer radius of the emitting region, the inclination angle of the disk and the efficiency of the disk emissivity. With this in mind, we modeled the broad Fe K$\alpha$ line with \texttt{XSPEC} model \texttt{laor} that produces an emission line from an accretion disk and includes general relativistic effects \citep{Laor91}. Keeping the emissivity as a function of radius ($\propto R^{-3}$), the outer radius fixed at 400 $GM/c^2$, and the inclination fixed at 30$^\circ$, we let the inner radius vary. 

\begin{figure*}[t]
\begin{center}
\includegraphics[angle=90,scale=.5]{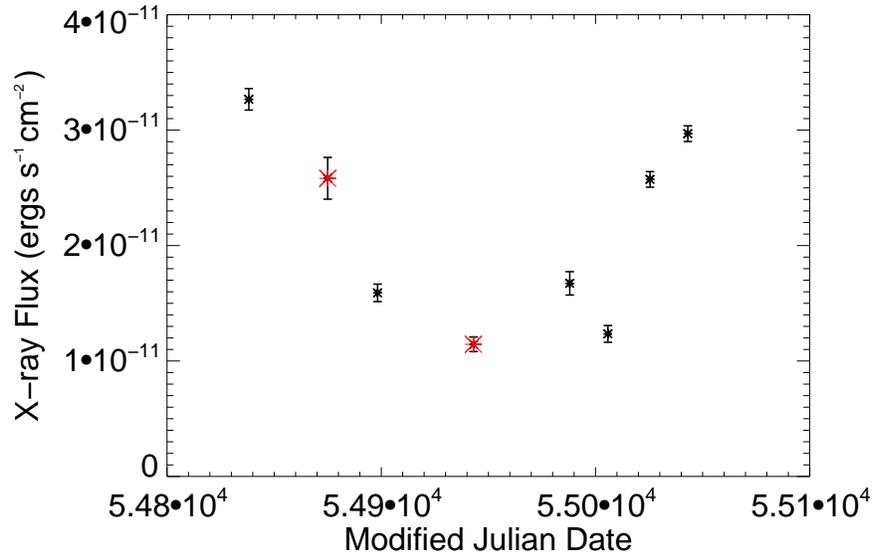}
\caption{\footnotesize{ This is the X-ray light curve taken by \emph{Chandra} in the continuous clocking mode between 2--10 keV. It shows a variability of a factor of 3. The points in red do not have simultaneous radio observations and are not used in the correlation analysis.}}
\label{xraylightcurve}
\end{center}
\end{figure*}

Finally, in order to compare to the fundamental plane of accretion onto black holes as described by \cite{Gultekin09}, just the power-law  component within the energy range of 2--10 keV was used to compute the total flux associated with each observation. This component contributes to more than 95\% of the total flux in the 2--10keV range, with the broad Fe K$\alpha$ line contributing 2--5\% of the total flux. We plot this light curve in Figure \ref{xraylightcurve}, and list the results for each spectrum in Table \ref{xrayobs}. The continuum flux varies by almost a factor of 3, which can be attributed to fluctuations in mass accretion rate. 

\subsection{Radio}
A combination of the Very Large Array (VLA) and Extended Very Large Array (EVLA) antennae were used to observe NGC 4051 six times between 31 Dec 2008 (MJD 54831.3) and 31 July 2009 (MJD 55043.1). These observations were approximately 1 hour integrations centered at 8.4 GHz with a bandwidth of 50 MHz in two channels with two polarizations. We chose this frequency to ensure self-absorption of the synchrotron emission was not a problem. Exposure dates and times are listed in Table \ref{radioobs}. During each observation, the antennae switched from on source, located at 1h203m1s $+44^\circ31'8''$, to a phase calibrator located at 12h21m4s $+44^\circ11'4''$ every 3.33 s.  3C 286 was used as the flux calibrator with 3.33 s of integration at the end of each observation. During the seven months spanning the observations, the VLA/EVLA evolved from its A configuration, which has its longest baseline of 36.4 km, to its C configuration with a maximum baseline of only 3.4 km. This significantly changed the resolution of the resulting images, and steps were taken to ensure a consistent comparison among the six observations. The resolution during each configuration is listed in Table \ref{radioobs}. 

\begin{figure*}[t]
\begin{center}
\includegraphics[scale=1.0, trim = 10mm 160mm 0mm 0mm, clip]{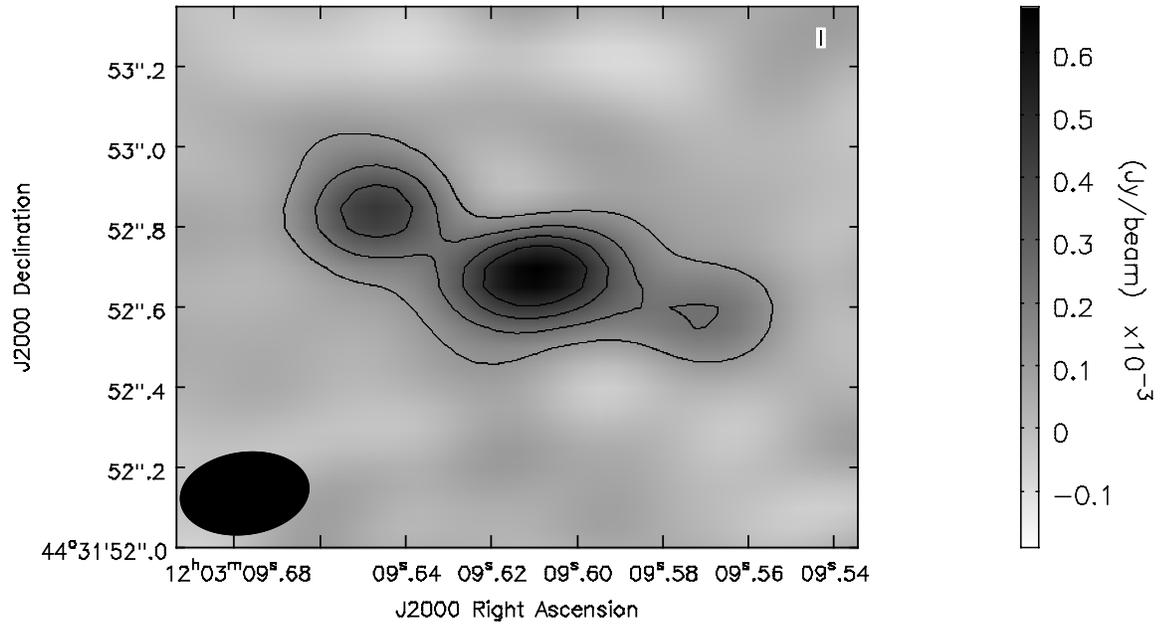}
\vspace{-3cm}
\end{center}
\caption{\footnotesize{Shown here is a radio observation made by VLA/EVLA in the A-configuration on 31 Dec 2009 (MJD 54831.3). The contours are [0.2, 0.4, 0.6, 0.8] $\times$ 0.6 mJy beam$^{-1}$. The beam pattern ($3'' \times 2''$) is also shown in the lower left corner. The center is associated with the black hole, while the extended lobes northwest and southeast are associated with the endpoints of jets of the system. }}
\label{fig3}
\end{figure*}

The Common Astronomy Software Applications (\texttt{CASA}) package, version 3.0.0, developed by National Radio Astronomy Observatory (NRAO) was used to reduce the radio observations. The routine \texttt{setjy} was used to set 3C 286 as the flux calibrator, while the routine \texttt{gaincal} set the gain calibration incorporating both the phase calibrator source located at 12h21m4s $+44^\circ1\arcmin4\arcsec$ and 3C 286 as references. Either antenna VA06, VA08, VA10, or VA12 was used as the reference antennae for these calibrations. Antennae that were off or showed discontinuities that were not characteristic of the overall data in either phase or amplitude were excluded, resulting in only one or two EVLA antennae per data set. The \texttt{CLEAN} algorithm was then run to create an image, using a threshold of 0.1 mJy as well as Briggs weighting and robust parameter of 0.5. Briggs weighting provides a smooth transition between natural and uniform visibility weighting and is characterized by a robust parameter where 2.0 is approximately natural weighting and -2.0 is approximately uniform \citep{Briggs95}. No self-calibration was done because the source was too faint, which is consistent with methods described by \cite{Giroletti09}, who also observed NGC 4051 at 8.4 GHz. Figure \ref{fig3} shows an example of the type of images that were produced. 

\begin{figure*}[t]
\begin{center}
\includegraphics[angle=90,scale=.5]{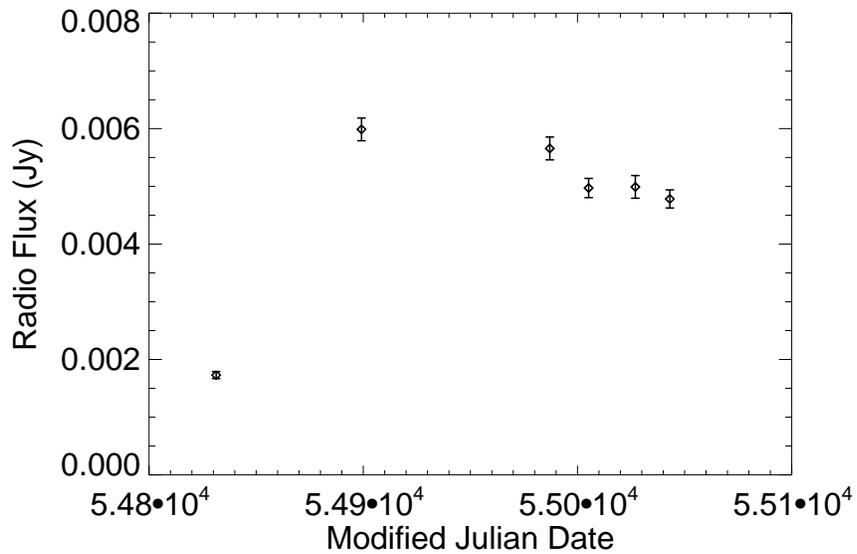}
\caption{\footnotesize{ This figure shows the 8.4 GHz light curve taken by the VLA/EVLA. The radio flux density varies by more than a factor of  3. }}
\label{radiolightcurve}
\end{center}
\end{figure*}

After the \texttt{CLEAN} algorithm was run, the flux density was measured using \texttt{CASA}'s \texttt{imfit}, which fit Gaussian curves to the peak intensities. Because of the four different configurations, we were only able to resolve structures in the A configuration, as seen in Figure \ref{fig3}, which afforded the best resolution. In order to make a fair comparison between all the measurements, the Gaussian fits were restricted to an area of approximately 8$'' \times 6''$, ensuring the inclusion of all the structures present in the images. Fortunately, as seen in the work done by \cite{Giroletti09}, there does not appear to be any extensive large scale diffuse emission that would be resolved out in higher resolution configurations. If this were the case, the flux would decrease with higher resolution, when in fact Table \ref{radioobs} demonstrates that the flux in the B configuration, a higher resolution, is larger than the flux in the C configuration, a lower resolution. This strongly suggests that the radio variability observed between the B and C configurations is intrinsic, not instrumental. However, it is unclear if the diffuse emission has been resolved out in the A configuration. For this reason, we quote our results with and without the A array measurements. The measurements including the integral flux densities as well as the peak flux densities are listed in Table \ref{radioobs} and the light curve is shown in Figure \ref{radiolightcurve}. The errors in these measurements also include a systematic error of 3\% of the total flux, added in quadrature, to account for the calibration errors. These measurements do vary by a factor of 3, which is comparable to the X-ray variations. In order to compare to \cite{Gultekin09}, we scaled the fluxes as $F_\nu \propto \nu^{-0.5}$ from 8.4 GHz to 5 GHz as \cite{Ho02} did.

Inspection of the image produced in the A configuration reveals evidence of jet production. As shown in Figure \ref{fig3}, three structures are present: a central source and two extended lobes. The central radio lobe is likely associated with the black hole, while the northwest and southeast lobes may be associated with the endpoints of the outflows in this system. The alignment and varying brightness between the three lobes suggests this is a jet structure where the jet is partially projected into our line-of-sight, placing the northwest lobe closer and less obscured then the southeast lobe. We also note that these lobes are not seen in the antennae psf and are therefore not artifacts of the \texttt{Clean} algorithm. Observations by \cite{Christopoulou97}, \cite{Giroletti09}, and \cite{Kukula95} note similar structures as well as the potential evidence for jet production. 

\subsection{Comparison}

These measurements by themselves can give insights into the physical processes that occur in AGN, but through a comparison of the X-ray and radio luminosities we can explore how accretion is coupled to jet production. To quantify any correlation between the X-ray and radio fluxes, we calculated the Spearman's rank coefficient ($\rho$). This $\rho$ describes how well a data set exhibits a monotonic behavior between two variables; $\rho < 0$ corresponds to variables that are anti-correlated. We find a $\rho$ of $-0.66$, with a probability of only 16\% of zero correlation. 

\begin{figure*}[t]
\begin{center}
\includegraphics[scale=0.70, angle=90]{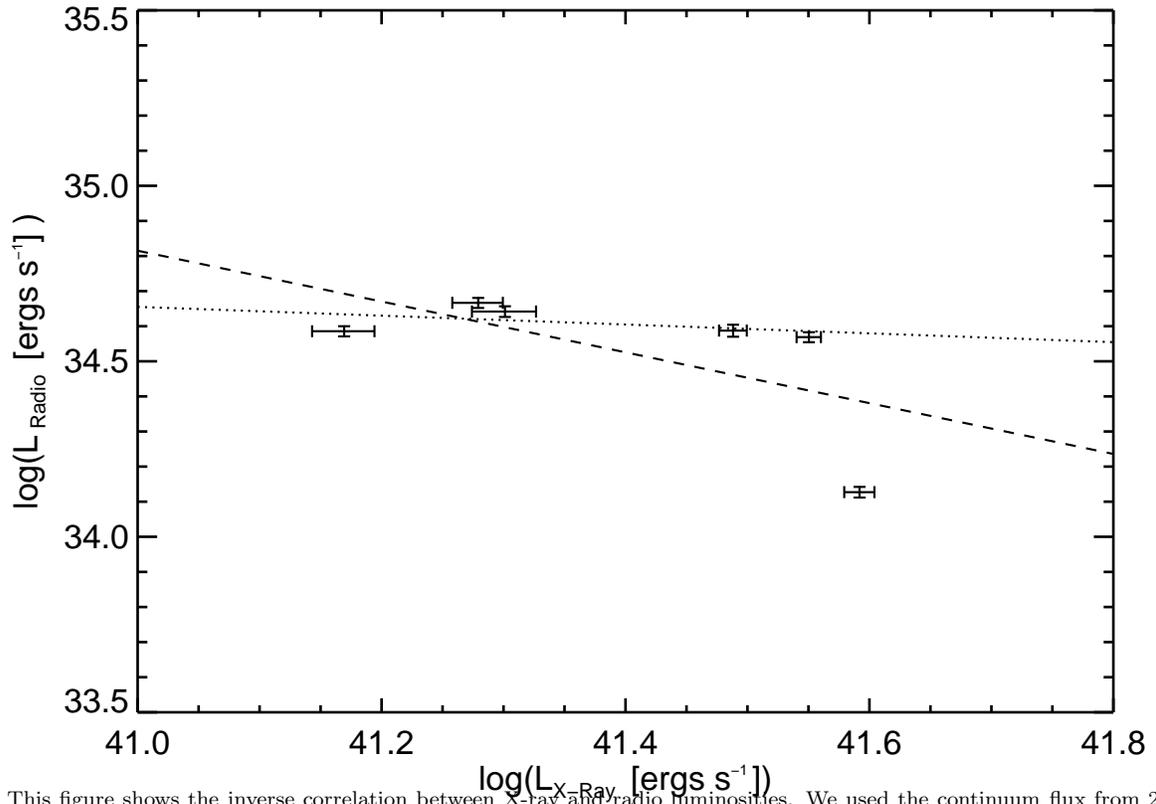}
\vspace{-1cm}
\end{center}
\caption{\footnotesize{This figure shows the inverse correlation between X-ray and radio luminosities. We used the continuum flux from 2--10 keV, and we scaled the radio flux as $F_\nu \propto \nu^{-0.5}$ \citep{Ho02} from 8.4 GHz to 5 GHz in order to correctly compare to \cite{Gultekin09}. The dashed line is a fit to our data given by the relation, $\log L_\mathrm{radio} = (-0.72 \pm 0.04) \log L_\mathrm{X-ray} + (64 \pm 2)$. The dotted line excludes the A configuration data point, and is described by the relation, $\log L_\mathrm{radio} = (-0.12 \pm 0.05) \log L_\mathrm{X-ray} + (40 \pm 2)$ }.}
\label{fig4}
\end{figure*}

Figure \ref{fig4} plots the X-ray vs. radio luminosities. We assumed a distance of 10.0 Mpc derived from the redshift, $z=0.002336$ \citep{Peterson04}, and cosmological parameter $H_\circ$=70 km $ s^{-1}$  Mpc to obtain these luminosities.  Figure \ref{fig4} also includes the least-squares fit of a first order polynomial to the data points, described by the following relation: 
\begin{equation}
\log L_\mathrm{radio} = (-0.72 \pm 0.04) \log L_\mathrm{X-ray} + (64 \pm 2)
\end{equation}
To test the robustness of the anti-correlation, we used an F-test to compare the results of the first order polynomial given above to fits with a zeroth order polynomial as well as the fundamental plane relation of $L_\mathrm{radio} \propto L_\mathrm{X-ray}^{0.64}$ . At an 81\% confidence level, we were able to rule out the flat model. However, at a 96\% confidence level, the fundamental plane relation is excluded by our model. 

We also looked at the correlation between the data points excluding the A configuration, in the case that this array resolved out a substantial amount of flux. We found that the correlation was still inversely proportional, described as,
 \begin{equation}
 \log L_\mathrm{radio} = (-0.12 \pm 0.05) \log L_\mathrm{X-ray} + (40 \pm 2)
 \end{equation}
Using the same F-test, our second model may be consistent with being flat, for it only excludes a flat distribution at the 60\% confidence level. However, the positive fundamental plane relation was again excluded, this time at a 98\% confidence level. 

\begin{figure*}[t]
\begin{center}
\includegraphics[scale=0.70, angle=90]{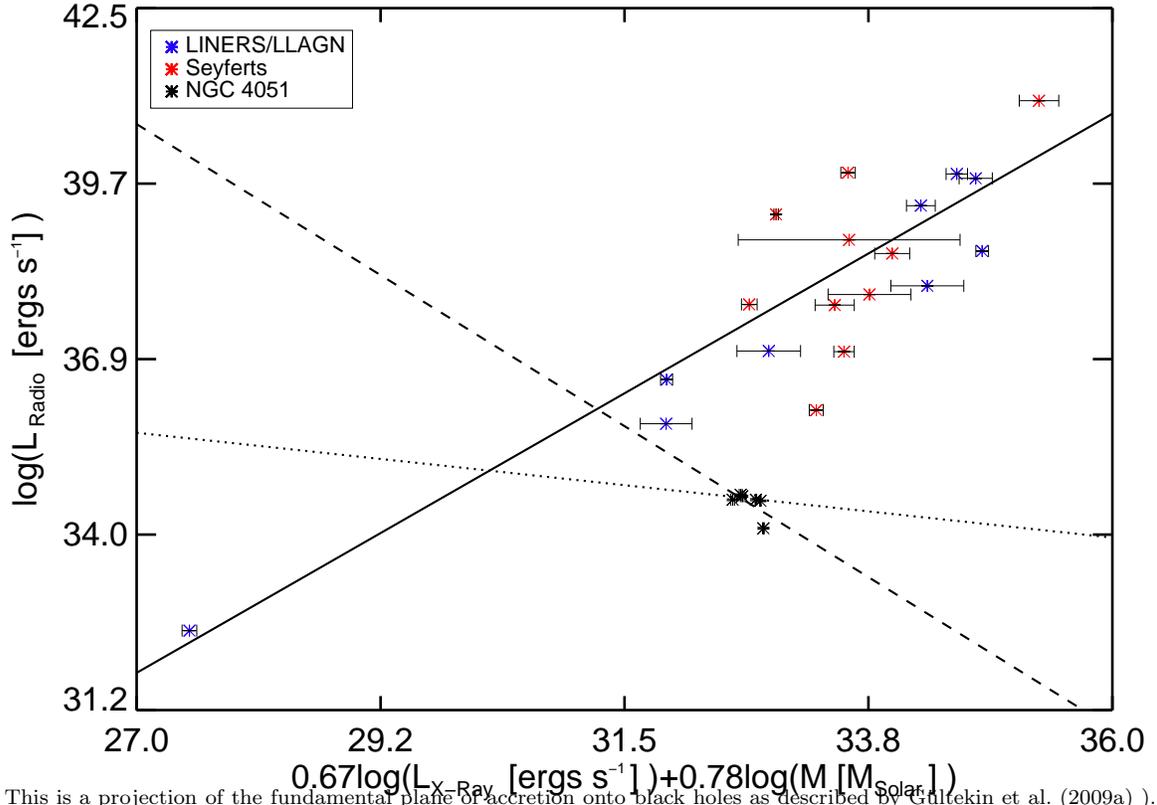}
\vspace{-1cm}
\end{center}
\caption{\footnotesize{This is a projection of the fundamental plane of accretion onto black holes as described by  \cite{Gultekin09} ). The red points are Seyfert galaxies, while blue points are LINERS or LLAGN. The solid black line is  \cite{Gultekin09} best fit line. Our data points are shown in black, with our best fit as the dashed line and dotted lines as described in Figure \ref{fig4}. }}
\label{fig5}
\end{figure*}

To place the data from NGC 4051 in context of the fundamental plane of accretion onto black holes, we plotted our data and best fit line against the data and the best fit line described in \cite{Gultekin09} in Figure \ref{fig5}. The best fit line found by \cite{Gultekin09} is described as follows,

\begin{equation}
\begin{split}
\log L_\mathrm{radio} = (0.67 \pm 0.12) \log L_\mathrm{X-ray} + \\
 (0.78 \pm 0.27) \log M_{BH} + (4.80 \pm 0.24)
\end{split}
\end{equation}

One should notice that our data are not described by the positive correlation of $L_\mathrm{radio} \propto L_\mathrm{X-ray}^{0.67\pm0.12}$ denoted by the solid line, but instead by the dashed line and dotted line described by  $L_\mathrm{radio} \propto L_\mathrm{X-ray}^{-0.72\pm0.04}$ and  $L_\mathrm{radio} \propto L_\mathrm{X-ray}^{-0.12\pm0.05}$, respectively. In fact, our relation differs from this relation by 11.6$\sigma$, while including the A configuration, and 6.5$\sigma$, while not including it, when dividing the difference between the power-laws by the larger of the two errors.  The fundamental plane is known to have a large scatter, and NGC 4051 lies close to the plane. This exercise, however, makes clear how different the X-ray and radio coupling in NGC 4051 appears to be.

\begin{figure*}[t]
\begin{center}
\includegraphics[angle=90,scale=.5]{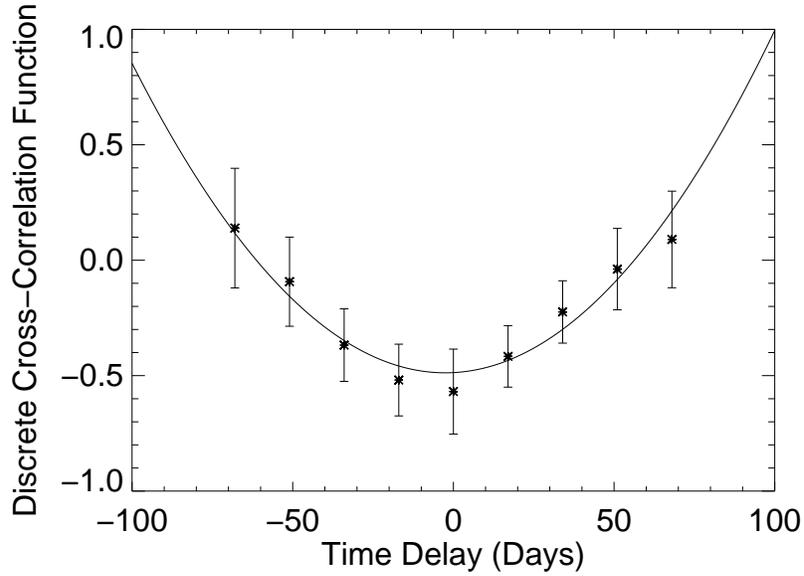}
\caption{\footnotesize{ This figure is of the discrete cross-correlation function (DCCF) between X-ray and radio luminosities. We fit the DCCF with a quadratic function, and the minimum occurs at -2.5$\pm$5.3 days with a DCCF value of -0.48$\pm$0.1. A negative delay implies that the radio is lagging the X-ray.}}
\label{DCCF}
\end{center}
\end{figure*}

Finally, we attempt to quantify whether or not there is a significant time delay between fluctuations in the X-ray and radio luminosities, which can be seen in Figures \ref{xraylightcurve} \& \ref{radiolightcurve} respectively. We used a discrete cross-correlation function (DCCF) as described by \cite{Edelson88} to quantify this delay. The sparse and uneven sampling of the all six observations poses a natural limitation on the results of this analysis. This led us to linearly interpolate the data within the seven month period with uniform spacing. The time steps between interpolated points were 17 days, corresponding to the shortest time between observations. \cite{White94} and \cite{Gaskell87} describe similar techniques for interpolating and cross-correlating data in their respective variability studies of AGN. The results, shown in Figure \ref{DCCF}, place the minimum of the DCCF of -0.48 $\pm$0.1 at -2.5$\pm$5.3 days, suggesting that X-ray dips are leading the radio flares.

\section{Discussion}
In this paper, we present eight \emph{Chandra} X-ray observations and six VLA/EVLA radio observations of Seyfert-1 NGC 4051 taken over a seven month period. By simultaneously measuring in X-ray and radio bands every two to four weeks apart, we are able to probe variations in the accretion rate that occur on the viscous timescale of a few days to weeks in NGC 4051. The observations reveal significant variability in both X-ray and radio bands. The first 8.4 GHz observation on MJD 54831.3 also shows evidence of jet production in the form of two distinct radio lobes, contrary to the idea that radio quiet galaxies like NGC 4051 lack jet production. In fact, work by \cite{Falcke01} and \cite{Nagar02} suggest that elongated, non-thermal emission is common in Seyfert and LLAGN. The lobes appear to be resolved into compact impact regions at very high resolution \citep{Giroletti09}. A variability analysis shows that the 2--10 keV X-ray luminosity and 8.4 GHz radio luminosity are inversely correlated according to $L_\mathrm{radio} \propto L_\mathrm{X-ray}^{-0.72 \pm 0.04}$.  This differs by $11.6 \sigma$ from the current fundamental plane relations which correlates the radio luminosity to X-ray luminosity as $L_\mathrm{radio} \propto L_\mathrm{X-ray}^{0.67 \pm 0.12}$ for a fixed mass \citep{Gultekin09}. Furthermore, if extended emission is resolved out in the highest resolution image, then excluding the A configuration data point, the correlation is $L_\mathrm{radio} \propto L_\mathrm{X-ray}^{-0.12 \pm 0.05}$. This still differs from the fundamental plane by more than 6$\sigma$ and is not consistent with the relation.

The results help to shed light on the fundamental plane of accretion onto black holes \citep{Merloni03, Falcke04, Gultekin09}. This plane demonstrates that accretion is linked to jet production, as evidenced by the positive correlation between X-ray and radio luminosity. The measurements in the fundamental plane are generally not simultaneous, and one goal of this study was to understand if simultaneity between X-ray and radio measurements would reduce the scatter in the plane. NGC 4051 was found to lie near to the fundamental plane, but showed a negative correlation between X-ray and radio luminosities, contrary to the positive correlation of the plane. The inverse correlation suggests that some systems may lie near to the fundamental plane, but vary like NGC 4051, moving across the plane and thus increasing the total scatter. 

At least a few other case studies have also shown a separate and inverse correlation deviating from the positive correlation suggested by the fundamental plane. The first is 3C 120, a SMBH with a mass of approximately 5.5 $\times 10^7 M_\odot$ that is observed to accrete at approximately 10\% $L_{Edd}$ \citep{Peterson04}. \cite{Chatterjee09} present a five year study of 3C 120 using RXTE and VLBI to obtain X-ray and radio data respectively. They use a discrete cross-correlation function to describe the correlation between the X-ray at 2.4--10 keV and the radio at 37 GHz. They find that the greatest amplitude in the DCCF to be $-0.68 \pm 0.11$, corresponding to an inverse correlation at a 90\% confidence level. This amplitude is consistent with the results of the cross-correlation analysis in NGC 4051. The minimum in of the DCCF of 3C 120 cited in \cite{Chatterjee09} corresponds to a time lag of 120 days $\pm 30$ days, with the X-ray dips leading leading the radio flares. This time delay corresponds to approximately 4 days in NGC 4051 when scaled using their respective masses, which is consistent with our data.

At the opposite end of the mass scale, the $14 M_\odot$ black hole GRS 1915+105 \citep{Greiner01} sometimes also shows an inverse relation between simultaneous X-ray and radio observations. \cite{Rau03} present a survey of X-ray and radio observations using RXTE and the Ryle Telescope between 1996 November to 2000 September. The 1--200 keV X-ray flux showed no correlation to the 15 GHz radio observations. However, the 20--200 keV continuum did show an inverse correlation to the 15 GHz flux described by a Spearman's rank-order correlation coefficient of -0.75. The hard X-ray band in GRS 1915+105 excludes direct emission from the disk, which is consistent with using X-ray emission instead UV emission in NGC 4051 and 3C 120.

The prototypical stellar-mass black hole, Cygnus X-1, may also show an inverse trend at high luminosity. \cite{Gallo03} study the disk-jet connections in this stellar-mass black hole by analyzing RXTE All Sky Monitoring X-ray data from 2--11 keV and the Ryle Telescope radio data at 15 GHz between 1996 January and January 2003. Cygnus X-1 does follow $L_\mathrm{Radio} \propto L_\mathrm{X-ray} ^{0.7}$, until it reaches approximately 2\% of its Eddington luminosity when the radio flux density turnovers. \cite{Gallo03} describe departures from this relation at high X-ray luminosity as quenching of its jet production as Cygnus X-1 moves into its ``high/soft" state. 

Observations of NGC 4051, 3C 120, GRS 1915+105, and Cygnus X-1 all show that X-ray and radio flux follow an inverse relation when observing at nearly simultaneous times when the sources are emitting at 1--10\% of Eddington. Given that an inverse correlation is seen in a quasar, a Seyfert, and two stellar-mass black holes at high Eddington fractions, it is possible that a distinct mode of disk-jet coupling holds at high Eddington fractions. This would go beyond a simple quenching of jet production, as discussed by \cite{Maccarone03} and \cite{Gallo03}, since the jet production does not turn off entirely \citep[evidenced by continuous radio emission and jet structures, especially in NGC 4051 and 3C 120;][]{Giroletti09, Chatterjee09}. This is the first study to probe a Seyfert galaxy in X-ray and radio bands on viscous timescales of its inner disk. In the future, we will undertake an extended monitoring campaign of NGC 4051 to further characterize this relation as well as determine any time lags between X-rays and radio fluxes.

\begin{acknowledgements}
The authors gratefully acknowledge Joan Wrobel, Phil Uttley, and Ian McHardy, and the anonymous referee for their insights and comments that improved this paper. They gratefully acknowledge Steven T. Myers for his help with the \texttt{CASA} software package.
J. M. M. gratefully acknowledges support through the \emph{Chandra} guest observer program.
S. M. gratefully acknowledges support from a Netherlands Organization for Scientific Research (NWO) Vidi Fellowship and from the European Community's Seventh Framework Programme (FP7/2007-2013) under grant agreement number ITN 215212 "Black Hole Universe".
E. M. C. gratefully acknowledges support provided by NASA through the \emph{Chandra} Fellowship Program.
\end{acknowledgements}

NOTE ADDED IN PROOF:  While our paper was being reviewed, a separate paper on radio and X-ray observations of NGC 4051 was accepted for publication \citep{Jones10}.  Our paper was accepted for publication only days later.  \cite{Jones10} report on many more radio observations, and employed RXTE to obtain X-ray flux points.  Treating the possibility of diffuse nuclear emission with great care, \cite{Jones10} find a nearly flat radio-X-ray correlation.  Our results are broadly consistent with that finding when the EVLA observation in the A configuration is excluded.  The main advantages of our work are the close timing of radio and X-ray observations, and the ability to separate distinct X-ray flux components in Chandra spectra.  Both papers show that NGC 4051 lies significantly below the Fundamental Plane.  Within the context of other black holes accreting at high Eddington fractions (a point of emphasis in this paper), both sets of results support the possibility that the coupling between the disk and jet in NGC 4051 may be different than in low-luminosity AGN and other extremely sub-Eddington sources. We thank Ian McHardy and Phil Uttley for helpful conversations regarding NGC 4051 and diffuse nuclear radio flux.

\begin{deluxetable*}{c c c c c c c c c}
\tablecolumns{8}
\tablewidth{0pt}
\tablecaption{X-ray Observations}
\tablehead{\colhead{Date of} & \colhead{Exposure} & \colhead{Count} & \colhead{$\Gamma$} & \colhead{$\Gamma$ Flux} & \colhead{ Disk Blackbody}& \colhead{Disk Blackbody} &\colhead{$\chi ^2 / \nu$}  \\ Observation & Time & Rate  & & (2--10keV) & Temperature & Flux (2--10keV) & & \\ (MJD)& (ks) & (counts s$^{-1}$) & & ($10^{-11} $ erg s$^{-1}$ cm$^{-2}$) & (keV) & ($10^{-15} $ ergs s$^{-1}$ cm$^{-2}$)  & }
\startdata 
54838.2 & 10.2 & 15.6 $\pm$ 0.04 & 2.24$ ^{+0.03 }_{-0.03 }$& 3.27$ ^{+0.09 }_{-0.09 }$   & 0.18$ ^{+0.02 }_{-0.02 }$ &10.7$^{+6.3}_{-3.7}$ &646.1/426 \\
$^\ast$ 54874.9 & 1.1 &16.3 $\pm$ 0.12 & 2.39$ ^{+0.11}_{-0.09 }$ & 2.58$ ^{+0.17 }_{-0.20 }$ & 0.17$ ^{+0.04 }_{-0.03 }$ & 6.1$^{+11.8}_{-4.1}$  & 244.6/213 \\
54898.3& 10.1 & 5.3 $\pm$ 0.02 &1.73$ ^{+0.08 }_{-0.03 }$& 1.59$ ^{+0.10 }_{-0.05 }$& 0.17$ ^{+0.01 }_{-0.02 }$& 4.3$^{+2.7}_{-1.0}$ & 394.0/381 \\
$^\ast$ 54943.1 & 10.1 & 4.7 $\pm$ 0.02 & 2.10$ ^{+0.05 }_{-0.07 }$ & 1.14$ ^{+ 0.05}_{-0.08 }$ & 0.18$ ^{+0.03 }_{-0.03 }$ & 3.7$^{+4.4}_{-1.4}$  &368.0/353 \\
54988.0 & 10.1 & 4.6 $\pm$ 0.02 & 1.33$ ^{+0.06 }_{-0.07 }$& 1.67$ ^{+0.09 }_{-0.12 }$ & 0.19$ ^{+0.01 }_{- 0.01}$ & 16.0 $^{+2.1}_{-3.1} $ &370.0/394 \\
55005.8 & 10.1 & 5.6 $\pm$ 0.02 & 2.01$ ^{+0.06 }_{-0.07 }$ & 1.23$ ^{+ 0.07}_{-0.08 }$ & 0.18$ ^{+0.01 }_{-0.01 }$& 8.1$^{+3.2}_{-2.4}$ &501.8/364 \\
55025.4 & 10.1 & 13.1 $\pm$ 0.04& 2.29$ ^{+0.03 }_{-0.03 }$& 2.57$ ^{+0.06 }_{-0.07 }$& 0.18$ ^{+0.02 }_{-0.02 }$& 5.1$^{+4.1}_{-2.2}  $ &578.7/411 \\
55043.1 & 10.1 & 14.9 $\pm$ 0.04 & 2.29$ ^{+0.03 }_{-0.02 }$& 2.97$ ^{+ 0.07}_{-0.06 }$& 0.17$ ^{+0.02 }_{-0.02 }$& 3.5$^{+2.3}_{-1.5}$  &627.7/420 \\
\enddata
\label{xrayobs}
\tablecomments{This table shows the X-ray observations made by \emph{Chandra} in the continuous clocking mode. The data was modeled using \texttt{Xspec} and unabsorbed fluxes for both the power-law and disk blackbody are presented here. The power-law used in the analysis varies by a factor of 3. The $^\ast$ refers to the observations that were not used in this analysis because of the lack of simultaneous radio measurements.}
\end{deluxetable*}

\begin{deluxetable*}{c || c c c | c c c c}
\tablecolumns{8}
\tablewidth{0pt}
\tablecaption{X-ray Fe K$\alpha$ line}
\tablehead{ & \multicolumn{3}{|c}{Narrow} & \multicolumn{4}{|c}{Broad} \\ \hline
\colhead{Date of} & \colhead{Fe K$\alpha$} & \colhead{Equivalent} & \colhead{Flux } & \colhead{Broad Fe K$\alpha$} & \colhead{Rin}  &\colhead{Equivalent} & \colhead{Flux} \\ Observation (MJD)& (keV) & Width (keV)  & ($10^{-13}$erg s$^{-1}$ cm$^{-2}$) & line (keV) & (GM c$^{-2}$) & Width (keV) & ($10^{-13}$erg s$^{-1}$ cm$^{-2}$)}
\startdata
54838.2 &  6.40$^{+ 0.04}_{- NA}$ & 0.04$ ^{+ 0.04}_{- 0.04}$ & 1.36$ ^{+1.29 }_{- 1.26 }$ & 6.95$ ^{+ NA }_{- 0.61 }$ & 8.2$ ^{+ 73}_{- 5}$ & 0.31$ ^{+ 0.24}_{- 0.18}$ & 7.07$ ^{+5.48 }_{- 4.11}$\\
$^\ast$ 54874.9  & -- & -- & --& 6.34$ ^{+ 0.42 }_{- NA }$ & 5.6$ ^{+ 20}_{- NA }$ & 0.54$ ^{+0.80 }_{- 0.49}$ & 10.1$ ^{+ 15.0}_{-9.18 }$\\
54898.3&  6.45$ ^{+ 0.09}_{-  0.07}$ & 0.04$ ^{+ 0.07}_{- NA}$& 0.84$ ^{+1.31 }_{-NA }$ & 6.62$ ^{+ NA}_{- 0.32 }$& 25$ ^{+ 165}_{- NA}$ & 0.15$ ^{+0.38 }_{-NA }$ & 2.27$ ^{+ 5.76}_{- NA }$\\
$^\ast$ 54943.1 &  6.40$ ^{+ 0.05}_{-  0.04}$& 0.14$ ^{+ 0.07}_{- 0.07}$& 1.81$ ^{+ 0.96 }_{- 0.98 }$ & 6.91$^{+ NA }_{- 0.56 }$ & 3.1$ ^{+ 13}_{- NA}$ & 0.70$ ^{+0.43}_{-0.45 }$ & 5.67$ ^{+3.53 }_{- 3.66}$\\
54988.0 &   6.39$ ^{+ 0.05}_{-  NA}$ & 0.06$ ^{+ 0.05}_{- 0.05}$ & 1.32$ ^{+ 1.20}_{- 1.25}$ & 6.35$ ^{+0.13 }_{- NA}$ & 1.2$ ^{+ 69}_{- NA}$ & 0.39$ ^{+0.26 }_{-0.28 }$ & 6.59$ ^{+4.52 }_{-4.75 }$\\
55005.8 &  6.36$ ^{+ 0.03}_{-  NA}$ & 0.12$ ^{+ 0.07}_{- 0.10 }$  & 1.73$ ^{+1.05 }_{- 1.44}$ & 6.34$ ^{+ 0.61}_{- NA}$ & 3.3$ ^{+ 31}_{- NA}$& 0.32$ ^{+ 0.26}_{-0.27 }$ & 3.94$ ^{+3.20 }_{-3.23 }$\\
55025.4 &  6.40$ ^{+ 0.11}_{-  NA}$ & 0.03$ ^{+ 0.04}_{- NA}$& 1.02$ ^{+1.31 }_{- NA}$ & 6.43$ ^{+ 0.13}_{- NA }$ & 44$ ^{+ 20}_{- 32}$ & 0.30$ ^{+0.12 }_{-0.11 }$& 6.60$ ^{+2.72}_{-2.49}$\\
55043.1 &  6.67$ ^{+ 0.07}_{-  0.08}$ &  0.02$ ^{+ 0.05}_{- 0.01 }$ & 0.64$ ^{+1.45 }_{-0.15}$& 6.34$ ^{+ 0.15 }_{- NA}$ & 20$ ^{+25 }_{-9 }$ & 0.29$ ^{+ 0.12}_{- 0.12}$ & 7.29$ ^{+2.98 }_{-2.86 }$\\
\enddata
\label{Feobs}
\tablecomments{This table shows the two Fe K$\alpha$ lines included in the X-ray models. To calculate the narrow Fe K$\alpha$ lines, \texttt{zgauss} with energy restricted between 6.35-6.97, a width of $\sigma=$0 (unresolved), redshift of z=0.002336 and varying normalization was used. For the broad Fe K$\alpha$ line, \texttt{laor} the energy was restricted to vary from 6.34-6.95 keV, the power-law dependence was frozen at 3, the inner edge was allowed to vary from 1.235 to 400 GM/c$^2$, the outer radius was frozen at 400 GM/c$^2$, the inclination angle was frozen at 30$^\circ$ and the normalization was allowed to vary. All the lines were at least marginally detected and therefore included in modeling techniques, except the narrow Fe K$\alpha$ on MJD 54874.9, where it registered 0 flux for the best fit parameters. The $^\ast$ refers to the observations that were not used in this analysis because of the lack of simultaneous radio measurements.}
\end{deluxetable*}

\begin{deluxetable*}{cc c c c c c}
\tablecolumns{7}
\tablewidth{0pc}
\tablecaption{Radio Observations}
\tablehead{\colhead{Date of Observation}  & \colhead{Exposure Time} & \colhead{Configuration} & \multicolumn{2}{c}{Resolution}  & \colhead{Flux Density} & \colhead{Peak Flux Density}\\ (MJD) & (ksec) & &(arcsec) & (pc) & (mJy) & (mJy)}
\startdata
54831.3 & 3.6 & A & 0.3 $\times$ 0.2 & 15 $\times$ 10 &  1.73 $\pm$ 0.06 & 0.65 $\pm 0.02$ \\ 
54899.2 & 3.5 & B &0.9 $\times$ 0.7 & 44 $\times$ 34 & 5.99 $\pm$ 0.20 & 1.13 $\pm  0.04$\\
54987.1 & 3.6 & BnC & 2.2 $\times$ 0.9 & 110 $\times$ 44 & 5.66 $\pm$  0.20 & 1.47 $\pm 0.05$ \\
55005.2 & 3.6 & C & 3.1 $\times$ 2.3 & 150 $\times$ 110 &  4.97 $\pm$  0.17 & 2.22 $\pm 0.08$\\
55027.0 & 2.4 & C & 2.5 $\times$ 2.0 & 120 $\times$ 97 &  4.99 $\pm$  0.20 & 2.08 $\pm 0.08$\\
55043.1 & 3.6 & C & 3.1 $\times$ 2.2 & 150 $\times$ 110 &  4.78 $\pm$  0.16 & 2.22 $\pm 0.08$\\
\enddata
\label{radioobs}
\tablecomments{ This table gives the radio observations taken at 8.4 GHz with a 50 MHz bandwidth in two channels. The VLA/EVLA was in 4 different configurations during the entire campaign, starting in the A with longest baseline and ending at C with the shortest baseline. In the analysis, we scaled the observations to 5 GHz as \cite{Ho02} did  using the total flux densities given in column 5. }
\end{deluxetable*}	


\clearpage

\bibliography{bib4051}

\begin{thebibliography}{46}
\expandafter\ifx\csname natexlab\endcsname\relax\def\natexlab#1{#1}\fi

\bibitem[{{Allen} {et~al.}(2006){Allen}, {Dunn}, {Fabian}, {Taylor}, \&
  {Reynolds}}]{Allen06}
{Allen}, S.~W., {Dunn}, R.~J.~H., {Fabian}, A.~C., {Taylor}, G.~B., \&
  {Reynolds}, C.~S. 2006, \mnras, 372, 21

\bibitem[{Blandford \& Konigl(1979)}]{Blandford79}
Blandford, R.~D., \& Konigl, A. 1979, ApJ, 232, 34

\bibitem[{{Briggs}(1995)}]{Briggs95}
{Briggs}, D.~S. 1995, in Bulletin of the American Astronomical Society,
  Vol.~27, Bulletin of the American Astronomical Society, 1444--+

\bibitem[{{Chatterjee} {et~al.}(2009){Chatterjee}, {Marscher}, {Jorstad},
  {Olmstead}, {McHardy}, {Aller}, {Aller}, {L{\"a}hteenm{\"a}ki}, {Tornikoski},
  {Hovatta}, {Marshall}, {Miller}, {Ryle}, {Chicka}, {Benker}, {Bottorff},
  {Brokofsky}, {Campbell}, {Chonis}, {Gaskell}, {Gaynullina}, {Grankin},
  {Hedrick}, {Ibrahimov}, {Klimek}, {Kruse}, {Masatoshi}, {Miller}, {Pan},
  {Petersen}, {Peterson}, {Shen}, {Strel'nikov}, {Tao}, {Watkins}, \&
  {Wheeler}}]{Chatterjee09}
{Chatterjee}, R., {et~al.} 2009, \apj, 704, 1689

\bibitem[{{Christopoulou} {et~al.}(1997){Christopoulou}, {Holloway}, {Steffen},
  {Mundell}, {Thean}, {Goudis}, {Meaburn}, \& {Pedlar}}]{Christopoulou97}
{Christopoulou}, P.~E., {Holloway}, A.~J., {Steffen}, W., {Mundell}, C.~G.,
  {Thean}, A.~H.~C., {Goudis}, C.~D., {Meaburn}, J., \& {Pedlar}, A. 1997,
  \mnras, 284, 385

\bibitem[{Cohen {et~al.}(1979)Cohen, Pearson, Readhead, Seielstad, Simon, \&
  Walker}]{Cohen79}
Cohen, M.~H., Pearson, T.~J., Readhead, A. C.~S., Seielstad, G.~A., Simon,
  R.~S., \& Walker, R.~C. 1979, ApJ, 231, 293

\bibitem[{{Crummy} {et~al.}(2006){Crummy}, {Fabian}, {Gallo}, \&
  {Ross}}]{Crummy06}
{Crummy}, J., {Fabian}, A.~C., {Gallo}, L., \& {Ross}, R.~R. 2006, \mnras, 365,
  1067

\bibitem[{{Denney} {et~al.}(2009){Denney}, {Watson}, {Peterson}, {Pogge},
  {Atlee}, {Bentz}, {Bird}, {Brokofsky}, {Comins}, {Dietrich}, {Doroshenko},
  {Eastman}, {Efimov}, {Gaskell}, {Hedrick}, {Klimanov}, {Klimek}, {Kruse},
  {Lamb}, {Leighly}, {Minezaki}, {Nazarov}, {Petersen}, {Peterson},
  {Poindexter}, {Schlesinger}, {Sakata}, {Sergeev}, {Tobin}, {Unterborn},
  {Vestergaard}, {Watkins}, \& {Yoshii}}]{Denney09}
{Denney}, K.~D., {et~al.} 2009, \apj, 702, 1353

\bibitem[{{Di Matteo} {et~al.}(2005){Di Matteo}, {Springel}, \&
  {Hernquist}}]{DiMatteo05}
{Di Matteo}, T., {Springel}, V., \& {Hernquist}, L. 2005, \nat, 433, 604

\bibitem[{{Edelson} \& {Krolik}(1988)}]{Edelson88}
{Edelson}, R.~A., \& {Krolik}, J.~H. 1988, \apj, 333, 646

\bibitem[{{Elvis} {et~al.}(1994){Elvis}, {Wilkes}, {McDowell}, {Green},
  {Bechtold}, {Willner}, {Oey}, {Polomski}, \& {Cutri}}]{Elvis94}
{Elvis}, M., {et~al.} 1994, \apjs, 95, 1

\bibitem[{{Fabian} {et~al.}(2002){Fabian}, {Celotti}, {Blundell}, {Kassim}, \&
  {Perley}}]{Fabian02}
{Fabian}, A.~C., {Celotti}, A., {Blundell}, K.~M., {Kassim}, N.~E., \&
  {Perley}, R.~A. 2002, \mnras, 331, 369

\bibitem[{{Fabian} {et~al.}(1989){Fabian}, {Rees}, {Stella}, \&
  {White}}]{Fabian89}
{Fabian}, A.~C., {Rees}, M.~J., {Stella}, L., \& {White}, N.~E. 1989, \mnras,
  238, 729

\bibitem[{Falcke {et~al.}(2004)Falcke, Kording, \& Markoff}]{Falcke04}
Falcke, H., Kording, E., \& Markoff, S. 2004, A\&A, 414, 895

\bibitem[{{Falcke} {et~al.}(2001){Falcke}, {Nagar}, {Wilson}, {Ho}, \&
  {Ulvestad}}]{Falcke01}
{Falcke}, H., {Nagar}, N.~M., {Wilson}, A.~S., {Ho}, L.~C., \& {Ulvestad},
  J.~S. 2001, in Black Holes in Binaries and Galactic Nuclei, ed. {L.~Kaper,
  E.~P.~J.~van den Heuvel, \& P.~A.~Woudt}, 218--+

\bibitem[{Ferrarese \& Merritt(2000)}]{Ferrarese00}
Ferrarese, L., \& Merritt, D. 2000, ApJ, 539, L1

\bibitem[{{Gallo} {et~al.}(2003){Gallo}, {Fender}, \& {Pooley}}]{Gallo03}
{Gallo}, E., {Fender}, R.~P., \& {Pooley}, G.~G. 2003, \mnras, 344, 60

\bibitem[{{Gaskell} \& {Peterson}(1987)}]{Gaskell87}
{Gaskell}, C.~M., \& {Peterson}, B.~M. 1987, \apjs, 65, 1

\bibitem[{Gebhardt {et~al.}(2000)Gebhardt, Bender, Bower, Dressler, Faber,
  Filippenko, Green, Grillmair, Ho, Kormendy, Lauer, Magorrian, Pinkney,
  Richstone, \& Tremaine}]{Gebhardt00}
Gebhardt, K., {et~al.} 2000, ApJ, 539, L13

\bibitem[{Gierlinski \& Done(2004)}]{Gierlinski04}
Gierlinski, M., \& Done, C. 2004, MNRAS, 349, L7

\bibitem[{Giroletti \& Panessa(2009)}]{Giroletti09}
Giroletti, M., \& Panessa, F. 2009, ApJ, 706, L260

\bibitem[{Greiner {et~al.}(2001)Greiner, Cuby, \& McCaughrean}]{Greiner01}
Greiner, J., Cuby, J.~G., \& McCaughrean, M.~J. 2001, Nature, 414, 522

\bibitem[{{G{\"u}ltekin} {et~al.}(2009{\natexlab{a}}){G{\"u}ltekin}, {Cackett},
  {Miller}, {Di Matteo}, {Markoff}, \& {Richstone}}]{Gultekin09}
{G{\"u}ltekin}, K., {Cackett}, E.~M., {Miller}, J.~M., {Di Matteo}, T.,
  {Markoff}, S., \& {Richstone}, D.~O. 2009{\natexlab{a}}, \apj, 706, 404

\bibitem[{{G{\"u}ltekin} {et~al.}(2009{\natexlab{b}}){G{\"u}ltekin},
  {Richstone}, {Gebhardt}, {Lauer}, {Tremaine}, {Aller}, {Bender}, {Dressler},
  {Faber}, {Filippenko}, {Green}, {Ho}, {Kormendy}, {Magorrian}, {Pinkney}, \&
  {Siopis}}]{Gultekin09b}
{G{\"u}ltekin}, K., {et~al.} 2009{\natexlab{b}}, \apj, 698, 198

\bibitem[{{Ho}(2002)}]{Ho02}
{Ho}, L.~C. 2002, \apj, 564, 120

\bibitem[{{Jones} {et~al.}(2010){Jones}, {McHardy}, {Moss}, {Seymour},
  {Breedt}, {Uttley}, {Kording}, \& {Tudose}}]{Jones10}
{Jones}, S., {McHardy}, I., {Moss}, D., {Seymour}, N., {Breedt}, E., {Uttley},
  P., {Kording}, E., \& {Tudose}, V. 2010, \mnras, accepted

\bibitem[{Jones {et~al.}(1974)Jones, O'dell, \& Stein}]{Jones74}
Jones, T., O'dell, S., \& Stein, W. 1974, ApJ, 188, 353

\bibitem[{Kalberla {et~al.}(2005)Kalberla, Burton, Hartmann, Arnal, Bajaja,
  Morras, \& P{\"{o}}ppel}]{Kalberla05}
Kalberla, P. M.~W., Burton, W.~B., Hartmann, D., Arnal, E.~M., Bajaja, E.,
  Morras, R., \& P{\"{o}}ppel, W. G.~L. 2005, A\&A, 440, 775

\bibitem[{{Kormendy} \& {Richstone}(1995)}]{Kormendy95}
{Kormendy}, J., \& {Richstone}, D. 1995, \araa, 33, 581

\bibitem[{Kukula {et~al.}(1995)Kukula, Pedlar, Baum, \& O'Dea}]{Kukula95}
Kukula, M.~J., Pedlar, A., Baum, S.~A., \& O'Dea, C.~P. 1995, MNRAS, 276, 1262

\bibitem[{Laor(1991)}]{Laor91}
Laor, A. 1991, ApJ, 376, 90L

\bibitem[{Maccarone {et~al.}(2003)Maccarone, Gallo, \& Fender}]{Maccarone03}
Maccarone, T., Gallo, E., \& Fender, R. 2003, MNRAS, 345, L19

\bibitem[{{Magorrian} {et~al.}(1998){Magorrian}, {Tremaine}, {Richstone},
  {Bender}, {Bower}, {Dressler}, {Faber}, {Gebhardt}, {Green}, {Grillmair},
  {Kormendy}, \& {Lauer}}]{Magorrian98}
{Magorrian}, J., {et~al.} 1998, \aj, 115, 2285

\bibitem[{Margon(1982)}]{Margon82}
Margon, B. 1982, Science, 215, 247

\bibitem[{{McHardy} {et~al.}(2004){McHardy}, {Papadakis}, {Uttley}, {Page}, \&
  {Mason}}]{McHardy04}
{McHardy}, I.~M., {Papadakis}, I.~E., {Uttley}, P., {Page}, M.~J., \& {Mason},
  K.~O. 2004, \mnras, 348, 783

\bibitem[{Merloni {et~al.}(2003)Merloni, Heinz, \& Matteo}]{Merloni03}
Merloni, A., Heinz, S., \& Matteo, T. 2003, MNRAS, 345, 1057

\bibitem[{{Miller}(2007)}]{Miller07}
{Miller}, J.~M. 2007, \araa, 45, 441

\bibitem[{{Mundt}(1985)}]{Mundt85}
{Mundt}, R. 1985, in Protostars and Planets II, ed. {D.~C.~Black \&
  M.~S.~Matthews}, 414--433

\bibitem[{{Nagar} {et~al.}(2002){Nagar}, {Falcke}, {Wilson}, \& {Ho}}]{Nagar02}
{Nagar}, N.~M., {Falcke}, H., {Wilson}, A.~S., \& {Ho}, L.~C. 2002, \nar, 46,
  225

\bibitem[{{Peterson} {et~al.}(2004){Peterson}, {Ferrarese}, {Gilbert}, {Kaspi},
  {Malkan}, {Maoz}, {Merritt}, {Netzer}, {Onken}, {Pogge}, {Vestergaard}, \&
  {Wandel}}]{Peterson04}
{Peterson}, B.~M., {et~al.} 2004, \apj, 613, 682

\bibitem[{Rau \& Greiner(2003)}]{Rau03}
Rau, A., \& Greiner, J. 2003, A\&A, 397, 711

\bibitem[{Reynolds(1997)}]{Reynolds97}
Reynolds, C. 1997, MNRAS, 286, 513

\bibitem[{{Richstone} {et~al.}(1998){Richstone}, {Ajhar}, {Bender}, {Bower},
  {Dressler}, {Faber}, {Filippenko}, {Gebhardt}, {Green}, {Ho}, {Kormendy},
  {Lauer}, {Magorrian}, \& {Tremaine}}]{Richstone98}
{Richstone}, D., {et~al.} 1998, \nat, 395, A14+

\bibitem[{Shakura \& Sunyaev(1973)}]{Shakura73}
Shakura, N., \& Sunyaev, R. 1973, A\&A, 24, 337

\bibitem[{{Uttley} {et~al.}(1999){Uttley}, {McHardy}, {Papadakis}, {Guainazzi},
  \& {Fruscione}}]{Uttley99}
{Uttley}, P., {McHardy}, I.~M., {Papadakis}, I.~E., {Guainazzi}, M., \&
  {Fruscione}, A. 1999, \mnras, 307, L6

\bibitem[{{White} \& {Peterson}(1994)}]{White94}
{White}, R.~J., \& {Peterson}, B.~M. 1994, \pasp, 106, 879

\end{thebibliography}

\end{document}